\documentclass[twocolumn,apl,amsmath,amssymb,showpacs]{revtex4-1}
\usepackage{epsf}      
\usepackage{graphicx}
\usepackage{color}
\usepackage{soul}
\usepackage{gensymb}
\usepackage{sidecap}
\usepackage{amsmath}

\begin{document}

\title{Na$_{4}$Co$_{3}$(PO$_{4}$)$_{2}$P$_{2}$O$_{7}$/NC composite as a negative electrode for sodium-ion batteries}
\author{Pravin K. Dwivedi, Simranjot K. Sapra, Jayashree Pati, and Rajendra S. Dhaka}
\email{rsdhaka@physics.iitd.ac.in}
\affiliation{Department of Physics, Indian Institute of Technology Delhi, Hauz Khas, New Delhi-110016, India}

\begin{abstract}
In recent years, the mixed phosphates based polyanionic electrode materials have attracted great attention in sodium-ion batteries due to their structural stability during cycling and open framework for ion diffusion. Here, we report the electrochemical performance of Na$_{4}$Co$_{3}$(PO$_{4}$)$_{2}$P$_{2}$O$_{7}$/nitrogen doped carbon (NCPP/NC) composite as a negative electrode (anode) for sodium ion batteries in the working potential range of 0.01--3.0~V. It delivers a reversible discharge capacity of 250~mAhg$^{-1}$ at 0.5~C current rate, which corresponds to the insertion/extraction of four sodium ions. The rate capability study indicates the reversible mechanism and highly stable capacity (61 mAhg$^{-1}$) even at high rate up to 5~C as compared to pristine NCPP. The incorporation of the N doped carbon spheres in the composite is expected to enhance the electronic/ionic conductivity, which plays an important role in improving the performance and stability up to 400 cycles at 1~C rate. Intriguingly, the analysis of cyclic voltammetry data measured at different scan rates confirm the capacitive/diffusive controlled mechanism and the extracted diffusion coefficient is found to be around 10$^{-10}$ cm$^2$s$^{-1}$. Our results demonstrate that the NCPP/NC composite is also a potential candidate as anode in sodium-ion batteries due to its three dimensional framework, cost effectiveness, enhanced specific capacity as well as further possibility of improving the stability. 

\end{abstract}
\date{\today}
\maketitle

\section{\noindent ~Introduction}

The expeditious development of environmentally benign renewable energy systems has brought significant advancement in the utilization of the wind and solar resources \cite{GielenESR19}. However, these are intermittent in nature due to the variability in time, duration and location. Therefore, the essential component is to storage the energy for the reliable dependence to fulfil the requirement of the society. In this respect, lithium-ion batteries (LIBs) have achieved immense success with reference to energy and power density and lent a platform for researchers to play-on across the globe, with the application in the grids, electric vehicles and portable electronic devices \cite{DunnSc11, LiAM18}. However, the non-uniform abundance and limited availability of the lithium resources \cite{TarasconNC10} obstructs to meet the increasing demand of the modern generation \cite{GrosjeanRSER12}. In order to search cost effective alternative, sodium-ion batteries (SIBs) are emerging as a pragmatic option owing to the large availability, uniform geographical distribution of the sodium resources in the earth's crust and also the SIBs follows the similar intercalation chemistry as LIBs \cite{SawwickiRSCAdv2015}. However, there are many challenges to develop electrode materials to accommodate the larger size sodium-ion, and the size also plays a crucial role in diffusion through electrolyte during charge/discharge. In comparison to the widely investigated layered \cite{Yuchemcomm2014,MaheshCI18, RakeshACSO19, MaheshEA20} as well as polyanionic type materials as cathode \cite{ChiharaJPS2013, Fangnanoletters2014, Dengnanoscale2015, Simran21}, search for the appropriate anode materials require immediate attention because the graphite anode cannot accommodate Na$^{+}$ ions at low-potential when used with Na salt electrolyte in sodium-ion batteries. Therefore, various carbonaceous materials have been investigated for their electrochemical performance towards SIBs as anode \cite{KomabaAFM2011, XiePE20, Xiao21}. However, their performance in terms of stability and storage capability is relegated than that of the corresponding lithium counterparts mainly due to the large radius of sodium. This leads to the reduced diffusion of sodium ions and huge volume change during the intercalation/deintercalation mechanism, hence, resulting in the severe structural degradation of the host material and poor cycleability and rate capability, which are not favorable for the application in the energy storage systems \cite{SaravanamAEM2013}. Hence, major consideration is imperative to find the suitable anode materials with excellent storage capacity and  rate capability as well as long cycle life.

In order to achieve excellent electrochemical performance, materials possessing open and stable three-dimensional (3D) frameworks with the larger diffusion pathways and interstitial space are expected to overcome the above challenges found in SIBs \cite{Simran21, SaravanamAEM2013, JianAEM2013}. In this context, the polyanionic compounds, owing to their 3D framework, structural and thermal stability are considered potential candidates \cite{Simran21}. For example, the mixed phosphates with general formula Na$_{4}$M$_{3}$(PO$_{4}$)$_{2}$P$_{2}$O$_{7}$ (M=Fe, Mn, Co, Ni) have attracted great attention as new class of electrode materials due to the low volume expansion arising from the strongly covalent bonded framework with the excellent cycling performance and longer lifetime of the battery system \cite{SanzCM2001, KangSCE20, ZarrabeitiaCM19, NoseEC13, NoseJPS13, RameshRSCA20, SanzSSC1996}. These mixed phosphate materials have three dimensional orthorhombic framework containing MO$_{6}$ octahedra and PO$_{4}$ tetrahedra sharing corners with pyrophosphate group via P--O--P interlayer linkage \cite{MoriwakeJPS2016}. Interestingly, the sodium ion occupies four sites in the three dimensional (3D) diffusion pathways. In comparison to the ion diffusivity in 1D ion channel, the 3D ion channels have greater advantage due to the availability of the alternative diffusion pathways, which exist even if defects in the bulk hinders the ion mobility in one particular direction of 3D ion channel \cite{ZarrabeitiaCM19, NoseEC13, NoseJPS13, RameshRSCA20, SanzSSC1996, MoriwakeJPS2016, LIUJMCA19, ZarrabeitiaEA21}. More importantly, due to the availability of large 3D channels in the Na$_{4}$Co$_{3}$(PO$_{4}$)$_{2}$P$_{2}$O$_{7}$ (NCPP), it can be considered as a good 3D ionic conductor \cite{ZarrabeitiaCM19, NoseEC13, NoseJPS13, RameshRSCA20, SanzSSC1996, MoriwakeJPS2016, LIUJMCA19, ZarrabeitiaEA21}. It also shows low-volume expansion during the cycling and has the high theoretical capacity (170~mAhg$^{-1}$). However, the lower energy densities and reversible capacities are observed due to the introduction of the P$_{2}$O$_{7}$ groups, leading to the reduced electronic conductivity. In order to enhance the electronic conductivity and reversible capacity, incorporation of carbon matrix in the structure is very effective. Moreover, the dopant heteroatoms (N, B, S, P) can provide pseudo-capacitance, enhance the interlayer distance, as well as generate extrinsic defects to provide more active sites for better Li-/Na-ion storage \cite{DwivediNanoscale2020, WangACSAMI14, KimNR19, ChenSm18, SunEA20}. In fact, N doping introduces donor states near the Fermi level, which can enhance  conductivity and reactivity by making the impurity sites active both electronically and chemically \cite{NevidomskyyPRL03, LatilPRl04}. Therefore, incorporation of N doped carbon matrix in NCPP electrode in the form of composite (NCPP/NC) is expected to increase the conductivity, and help in improving the energy density, rate capacity and cycle life of the Na-ion batteries \cite{KimNR19, ChenSm18, SunEA20}. 

As discussed above, recently the NCPP materials have been investigated in the high potential range as cathode \cite{ZarrabeitiaCM19, NoseEC13, NoseJPS13, RameshRSCA20, SanzSSC1996, MoriwakeJPS2016, LIUJMCA19, ZarrabeitiaEA21}; however, to the best of our knowledge, its electrochemical investigation has not been explored so far in the lower working potential as anode in sodium-ion batteries. Therefore, in this paper we report the synthesis of a composite of N-doped carbon and NCPP via a facile sol-gel route and examine its electrochemical performance as a negative electrode (anode) for SIBs and compared with the pristine NCPP material. We demonstrate a significant increase in the storage capacity of 250~mAhg$^{-1}$ at 0.5~C current rate in the working potential range of 0.01--3.0~V. The N-doped carbon incorporated on NCPP with the formation of a composite through the micro-sized spherical particles, provides fast electron transfer and provides sufficient bonding sites with high sodium ion diffusion in the range of 10$^{-10}$ cm$^2$s$^{-1}$. We found that a well designed NCPP/NC composite as anode shows excellent rate capability (61 mAhg$^{-1}$ at 5~C rate) and outstanding cycling stability (400 cycles) as compared to the pristine NCPP. Our study demonstrate the superior performance of the composite as anode for high performance sodium-ion  rechargeable batteries.

\section{\noindent ~Experimental}

 \textbf{Synthesis of N doped carbon spheres:} In order to prepare the N doped carbon spheres, we use 0.1 M of sucrose as a carbon precursor and 0.2 M of ammonium sulphate for N doping, which were dissolved in 100 ml of de-ionized water and stirred for 1~h. Afterwards, the solution mixture was transferred into the Teflon lined stainless steel hydrothermal reactor and heat treated for 4~h at 200$^{o}$C. This reaction mixture was filtered and washed with DI water and ethanol followed by overnight drying to obtained hydro-thermal char, which were annealed at 900\degree C for 3~h in an inert Ar atmosphere, which resulted in the N doped carbon spheres \cite{DwivediNanoscale2020, KimNR19, ChenSm18}. 
 
\begin{figure}
\centering
\includegraphics[width=3.4in]{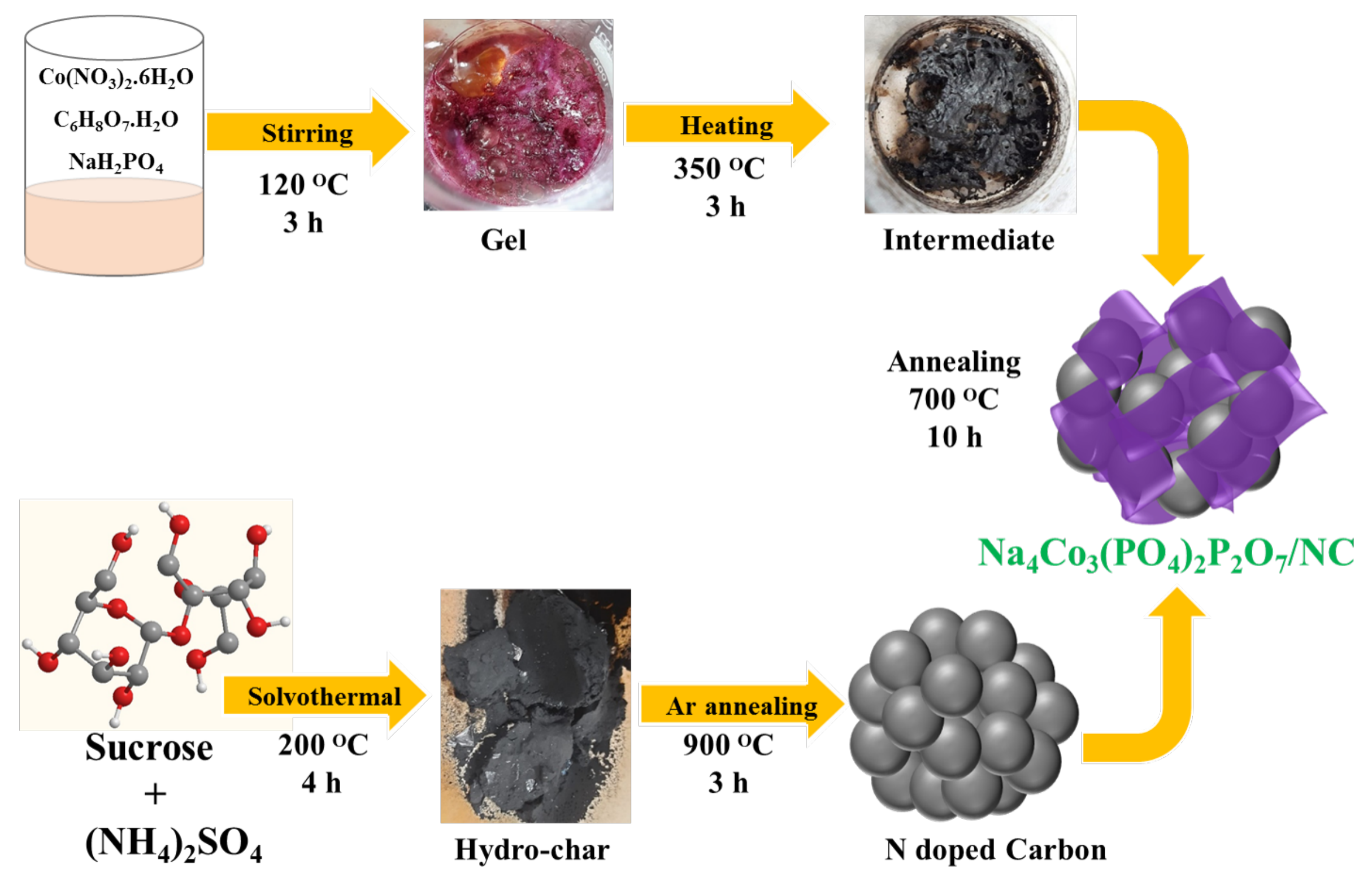}
\caption {Schematic illustration of synthesis process of the NCPP/NC composite material.}  
\label{Schematic1.png}
\end{figure}

\textbf{Synthesis of NCPP/NC nano-composite:}
Facile sol-gel process was used to synthesize the composites of NCPP/NC. First, the stoichiometric amounts of NaH$_{2}$PO$_{4}$.2H$_{2}$O, Co(NO$_{3}$)$_{2}$.6H$_{2}$O, and C$_{6}$H$_{8}$O$_{7}$.H$_{2}$O were added to de-ionized water in sequence with continuous stirring for 3~h at 120$^{o}$C. A moderate amount of citric acid was used as a carbon precursor and also as a chelating agent. The formed gel kept for calcination at 300$^{o}$C for 3~h in air atmosphere. The obtained product was grind into fine powder followed by mixing with 30\% of N doped carbon and grind again for 2-3~h by using mortar pestle. Further, an intermediate product was heated to 700$^{o}$C with 2$^{o}$C min$^{-1}$ rate and kept there for 10~h in an Ar atmosphere to obtain the NCPP/NC composites. The pristine NCPP was synthesized for comparison  in similar way except incorporation of the NC matrix. The schematic illustration of synthesis procedure of NC matrix and the NCPP/NC composites is shown in Fig.~\ref{Schematic1.png}.

\textbf{Material characterization:} The room temperature powder X-ray diffraction (XRD) measurements were carried out using Cu-K$\alpha$ radiation (1.5406 \AA) in the 2$\theta$ range of 5--80\degree using Rigaku Mini Flex 600 diffractometer. The recorded XRD data were analysed by Rietveld refinement using FullProf package, where we fitted the background using linear interpolation between data points. The surface morphology and the elemental mapping of the as prepared materials have been investigated through the field emission scanning electron microscope (FE-SEM) using the JEOL JSM-7800F Prime system. The high resolution transmission electron microscopy (HRTEM) measurements have been done with Tecnai G2 20 system. The Raman spectra of prepared pellets were recorded with Renishaw invia confocal Raman microscope using 2400~lines/mm grating at wavelength of 532~nm with laser power of 1~mW on the sample.

\textbf{Coin-cell fabrication:} The electrochemical properties of NCPP/NC composites have been examined by two-electrode coin cell configuration. In order to fabricate the anode, the slurry was prepared by mixing the active material with the acetylene black and the binder (polyvinylidenedifluoride, PVDF) in the ratio of 7:2:1 and dissolved in a N-methyl pyrrolodinone (NMP) solvent. The prepared slurry was coated on to Cu foil, which serve as a current collector using film coater and doctor blade. Subsequently, the coated copper sheet is air dried for 12~h and vacuum dried at 110$^{o}$C overnight in order to completely remove the solvent. The circular disks of 12~mm were cut using a disc cutter and vacuum dried at 100$^{o}$C for about 5~h. The coin cells of type CR2032 were assembled in an Argon-filled glovebox (UniLab Pro SP from MBraun) having oxygen and water content less than 0.1~ppm. The half cells are assembled using sodium metal foil as a counter electrode, a glass microfiber paper as a separator (Glass Fiber Filter GB 100R) and the electrolyte as 1 M NaClO$_4$ in diethyl carbonate (DEC)--ethylene carbonate (EC) in 1:1 volume. 

\textbf{Electrochemical measurements:} The cyclic voltammetry (CV) measurements were performed using VMP-3 Biologic pontentiostat at different scan rates between 0.01-3.0~V for the anode. The galvanostatic charge-discharge measurements for the cells were carried out by using Neware battery analyser BTS400 as well as VMP-3 Biologic pontentiostat. The electrochemical impedance spectroscopy (EIS) data were collected in the frequency range from 100~kHz to 10~mHz with an AC amplitude of 10~mV using Biologic VMP-3 potentiostat.

\section{\noindent ~Results and discussion}

\begin{figure}
\includegraphics[width=3.55in]{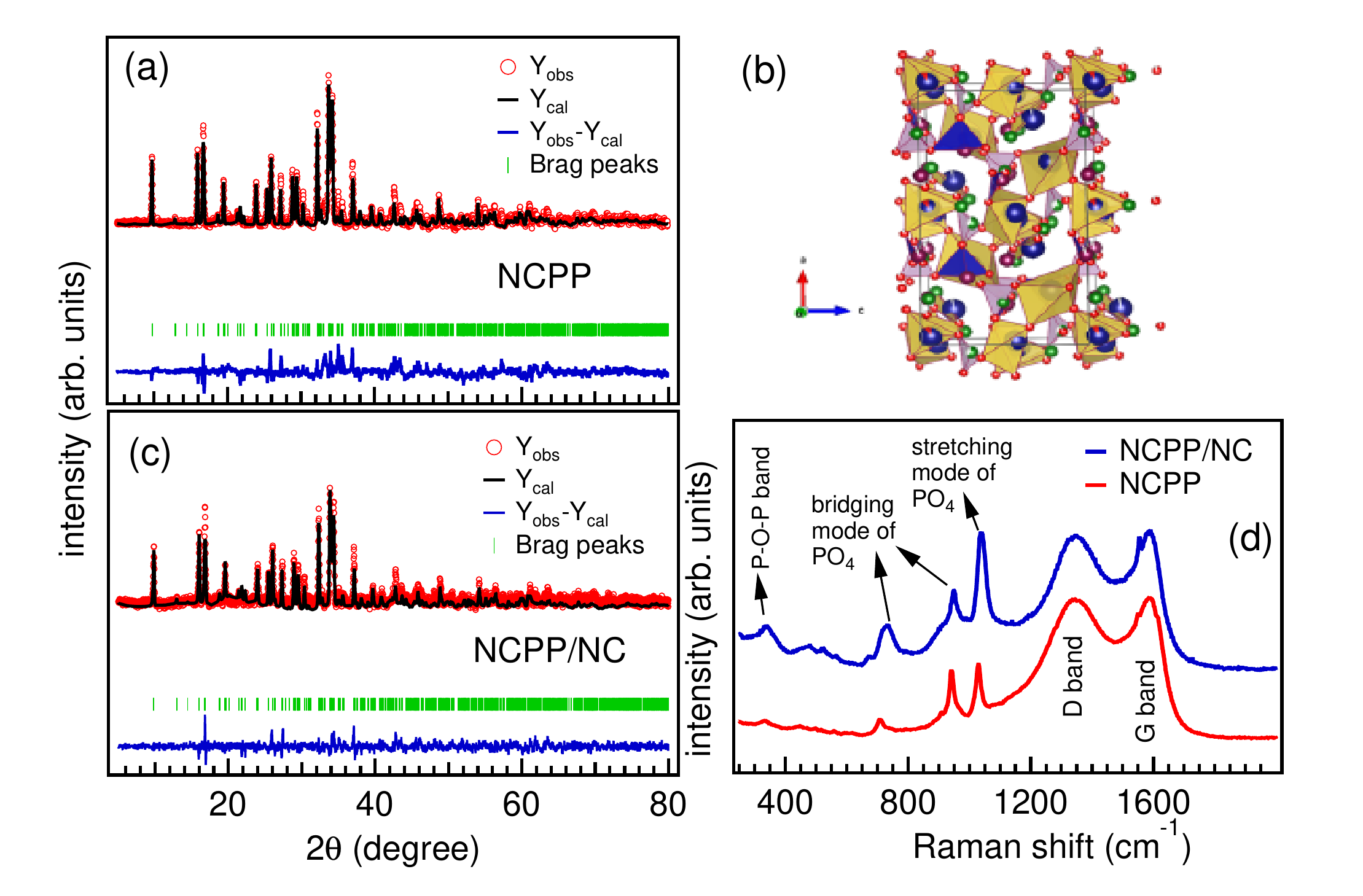}
\includegraphics[width=3.55in]{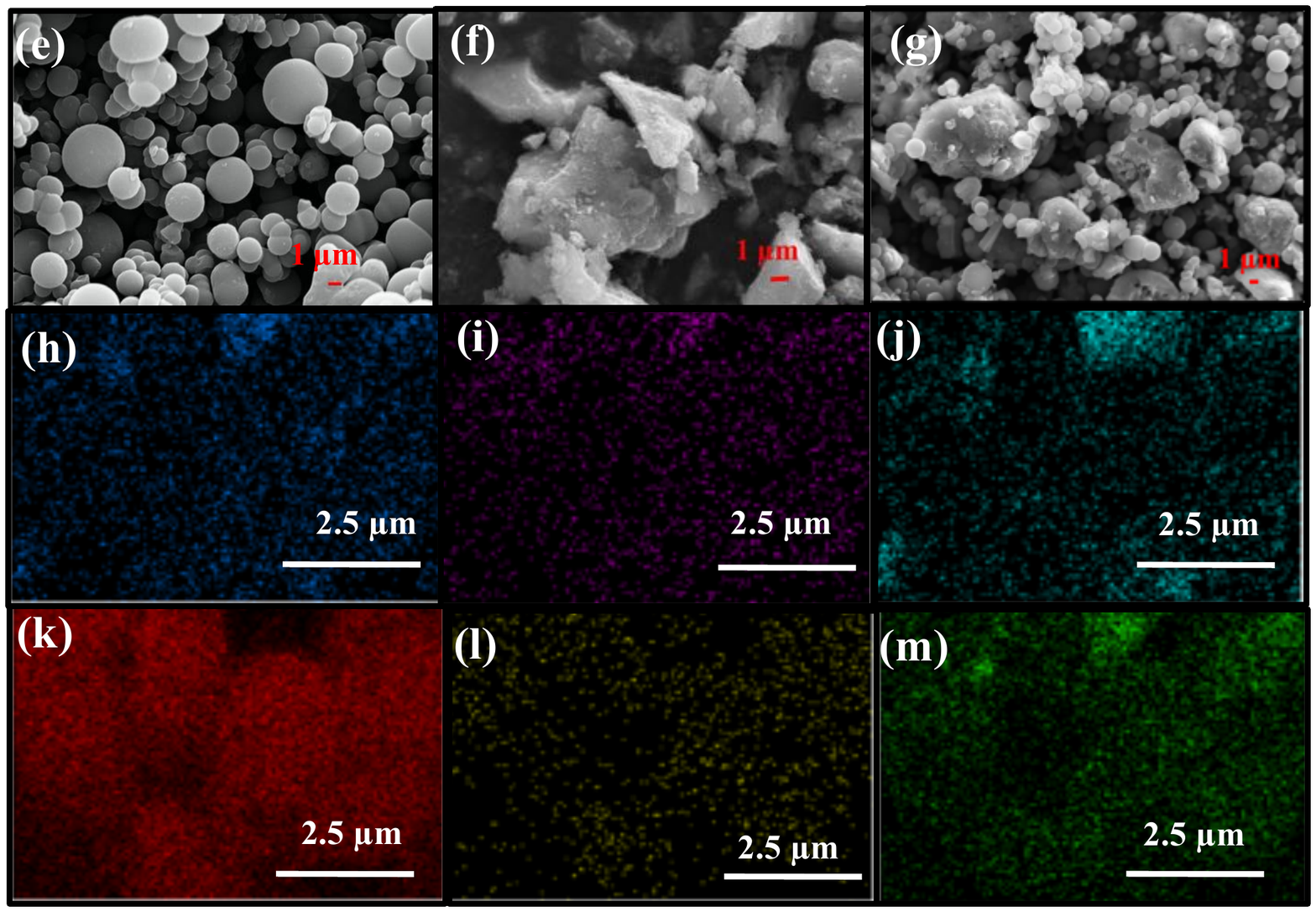}
\includegraphics[width=3.55in]{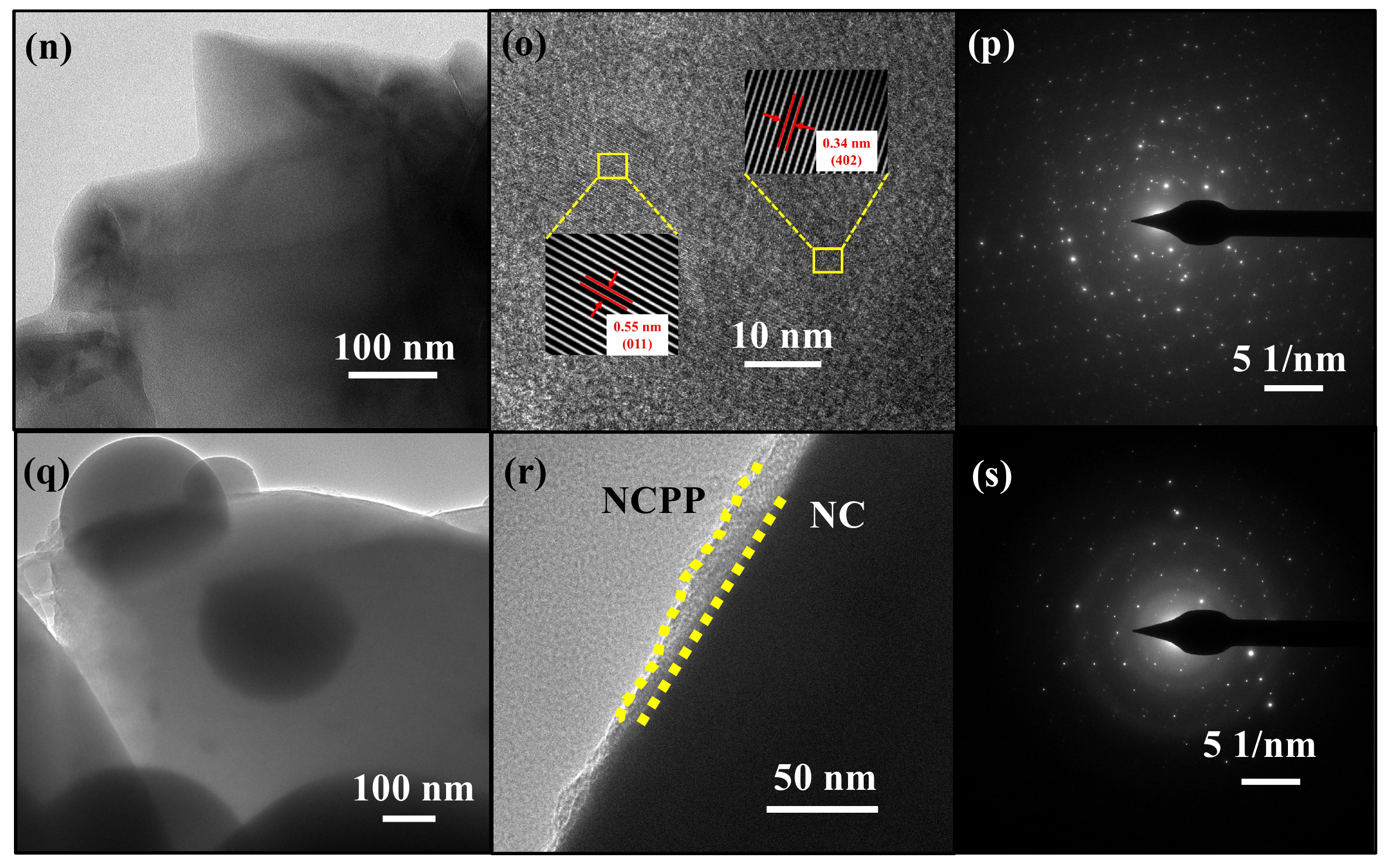}
\caption{The Rietveld refined XRD patterns of the (a) NCPP and (c) NCPP/NC samples; (b) the schematic illustration of NCPP crystal structure, where Co, P, Na and O are shown by the purple, green, blue ad red color respectively, visualised in the VESTA software; (d) the Raman spectra of the NCPP and NCPP/NC composite; the SEM images of the (e) N doped carbon spheres, (f) pristine NCPP sample, and (g) NCPP/NC composite; the elemental mapping images of (h-m) Na, Co, P, C, N and O, respectively in the NCPP/NC composite; the TEM images of (n) pristine NCPP, and (q) NCPP/NC composite; the high resolution TEM images of (o) pristine NCPP, and (r) NCPP/NC composite; the SAED patterns of (p) pristine NCPP, and (s) NCPP/NC composite samples.}
\label{characterization}
\end{figure}

The Rietveld refinement of the XRD patterns of NCPP and NCPP/NC are shown in Figs.~\ref{characterization}(a, c), respectively, which confirm the formation of single phase and mixed phosphates structure of both the samples. The lattice parameters of the NCPP/NC compound obtained are: $a=$ 18.0456 \AA, $b=$ 6.518 \AA, $c=$ 10.561 \AA, and a unit volume of 1242.13~\AA$^3$ corresponding to an orthorhombic Pn2$_1$a space group.The crystal structure of NCPP/NC is displayed in Fig.~\ref{characterization}(b) using the program VESTA. The NCPP consists of corner/edge-sharing of CoO$_6$ octahedra and PO$_4$ tetrahedra where the PO$_4$ units share one edge and two corners with the adjacent CoO$_6$ octahedra, forming a pseudo-layered structure along the $bc$-plane. These layers are successively bridged by P$_2$O$_7$ pyrophosphate units along the $a-$axis to form large 3D ion channels having four types of Na$^{+}$ ion sites (Na1, Na2, Na3, Na4). 
\begin{figure*}
\centering
\includegraphics[width=6.8in]{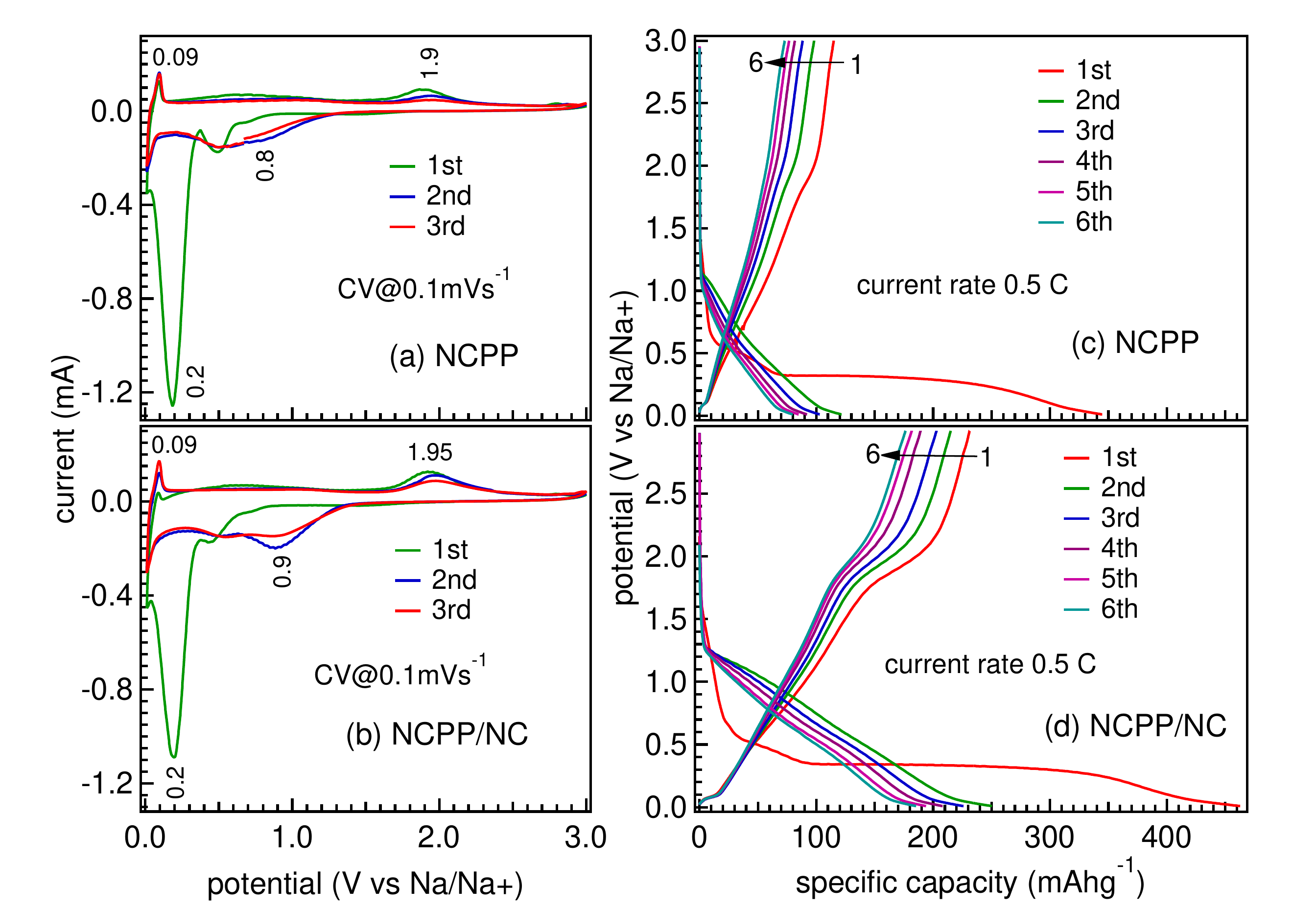}
\caption{(a, b) Cyclic voltammetry (CV) curves of the NCPP and NCPP/NC anodes, respectively, measured at 0.1 mVs$^{-1}$ scan rate, and  (c, d) Charge-discharge profiles of the NCPP and NCPP/NC anodes, respectively, measured at 0.5 C current rate.} 
\label{NCPP_NCPPNC_CV_C_DC}  
\end{figure*}
Further, we use Raman spectroscopy to check the structure, which confirms the presence of graphitized carbon as the spectra are shown in Fig.~\ref{characterization}(d). For the NCPP/NC sample, the characteristic peaks of D- and G-bands are observed at 1343 and 1588 cm$^{-1}$, respectively. Whereas, these peaks are at 1349 and 1595 cm$^{-1}$, respectively, for the NCPP sample. These are correspond to the disordered and graphitized carbons, which could be attributed to N doping in carbon and decomposed carbons \cite{DwivediNanoscale2020}. The peaks appear in the range of 300 cm$^{-1}$ to 1200 cm $^{-1}$ consistent with the internal modes of P$_{2}$O$_{7}$ \cite{SenthilkumarECA2015}. The diphosphate ion consists of two PO$_{4}$ tetrahedra with a central P-O-P linkage \cite{GangadharanJALCOM2002}. The Raman shift at 1037 cm$^{-1}$ corresponds to stretching mode of PO$_{4}$ group, peaks appear at 715 and 952 cm$^{-1}$ correspond to bridging mode of PO$_{4}$. The peak present at 343 cm$^{-1}$ indicates the deformation of P-O-P bond \cite{GangadharanJALCOM2002}. The detailed microstructures of the as-synthesized N doped carbon spheres, NCPP and NCPP/NC samples are characterized using FE-SEM and results are displayed in Figs.~\ref{characterization}(e--g), respectively. The spheres like morphology obtained for N doped carbon where as NCPP is composed of flakes and NCPP/NC is consist of flakes of NCPP and microspheres of N doped carbon. The size of the carbon spheres is in the range of 2 to 5~$\mu$m. The elemental mapping images confirm the homogeneous distribution of all the elements (Na, Co, P, C, N and O) in the sample, see Figs.~\ref{characterization}(h--m). To further investigate the detailed morphologies and structures of NC and NCPP/NC, we perform HR-TEM measurements, which demonstrate the sheet like structure of the NCPP sample, as shown in the Fig.~\ref{characterization}(n). In Fig.~\ref{characterization}(o) we show the high resolution image for the NCPP sample where the extracted interlayer distance was found to be 0.55~nm for (011) reflection and 0.34 for (402) reflection. Fig.~\ref{characterization}(q) shows the N doped carbon spheres surrounded by the NCPP sheet, and the high magnification image in Fig.~\ref{characterization}(r) indicates the surface boundary of NCPP and N doped carbon spheres. The selected area electron diffraction (SAED) patterns confirm the polycrystalline nature of both the NCPP and NCPP/NC samples, as shown in Figs.~\ref{characterization}(p, s), respectively. 

\begin{figure*}
\centering
\includegraphics[width=7.3in]{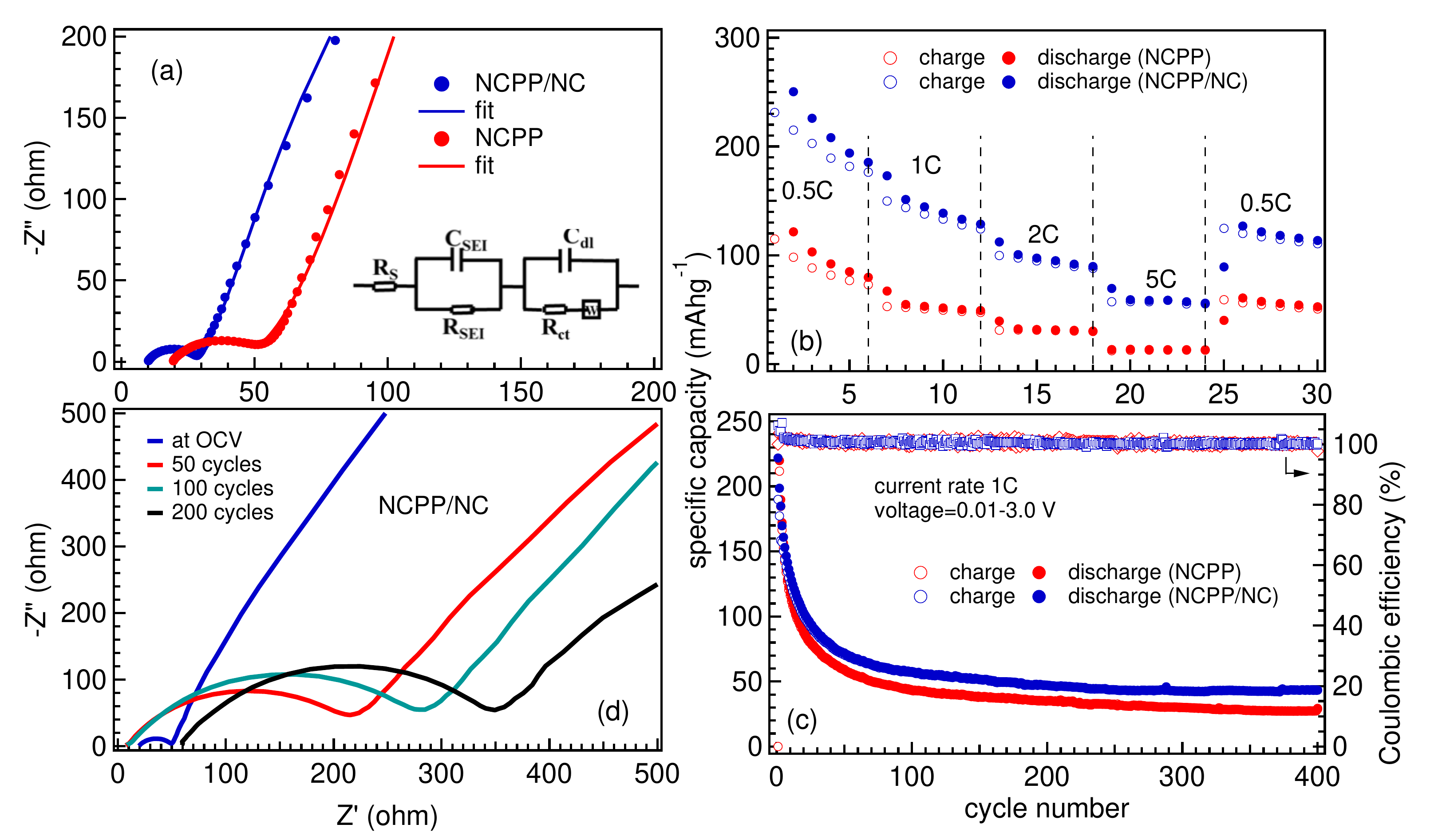}
\caption{(a) The Nyquist plot, (b) the rate capability, (c) the cyclic stability at 1~C current rate of the NCPP and NCPP/NC anodes, (d) electrochemical impedance spectra of the NCPP/NC anode during cycling at 1~C rate.}
\label{EIS_rate}  
\end{figure*}

Now we focus on to investigate the electrochemical performance of both the NCPP and NCPP/NC samples as negative electrode for sodium-ion batteries in the potential range of 0.01--3.0 V. The Cyclic voltametery (CV) measurements were performed vs. Na/Na$^{+}$ for first 3 cycles at a scan rate of 0.1~mVs$^{-1}$, as shown in Figs.~\ref{NCPP_NCPPNC_CV_C_DC}(a, b) for both the NCPP and NCPP/NC electrodes, respectively. The observed sharp cathodic peak at around 0.2~V is mostly due to the decomposition of the electrolyte and the film formation at the solid electrolyte interphase (SEI) in the first scan only as the peak disappears in subsequent cycles \cite{WangJMCA2015, PatraACS-AEM21}. The SEI is of crucial importance to the battery performance as it protects the anode by inhibiting the transfer of electrons from the anode to the electrolyte, while allowing only sodium ion diffusion. Moreover, the reversible peaks appear at 0.01~V in cathode region and 0.09~V in anodic region due to the intercalation and deintercalation of Na ions in carbon matrix \cite{WangJMCA2015, TangAEM2012}. From the second cycle onwards, the reversible peaks appear at around 0.9~V and 1.9~V due to the sodiation/desodiation process along with the electrochemical reduction/oxidation of Co in the active material \cite{DongNJC18, PalaniCS18, LiJMC13, PangJMCA19} of NCPP electrodes. Similar redox couples are observed in the CV curves of the pristine NCPP material, which suggests for the same electrochemical reactions. However, when compared to the NCPP/NC composite, the pristine NCPP material shows greater electrode polarization with the large voltage difference between cathodic and anodic peaks. The galvanostatic charge-discharge (GCD) measurements were carried out to find the specific capacity and the cycling stability of the NPCC/NC composite. The GCD profiles measured at 0.5~C rate in between 0.01 and 3.0~V voltage for both the NCPP and NCPP/NC anodes are shown in Figs.~\ref{NCPP_NCPPNC_CV_C_DC}(c) and (d), respectively. A prolonged voltage plateau appears below 0.5~V for both the electrodes due to the sodiation of NCPP and the formation of SEI layer. The slope in the range of 1.8 to 2.2~V during the charging represents the reversible oxidation of Co in the active material. The NCPP delivers the discharge capacity of 121~mAhg$^{-1}$ at second cycle, whereas the NCPP/NC composite exhibits 250~mAhg$^{-1}$. Interestingly, the NCPP/NC delivers significantly higher reversible capacity because of the the addition of the N doped carbon content in the composite, which improves the  electronic conductivity and increases the Na storage capacity. It also act as a buffer to sustain mechanical structure of electrode due to the volumetric expansion during sodiation/desodiation.

\begin{figure*}
\centering
\includegraphics[width=7.1in]{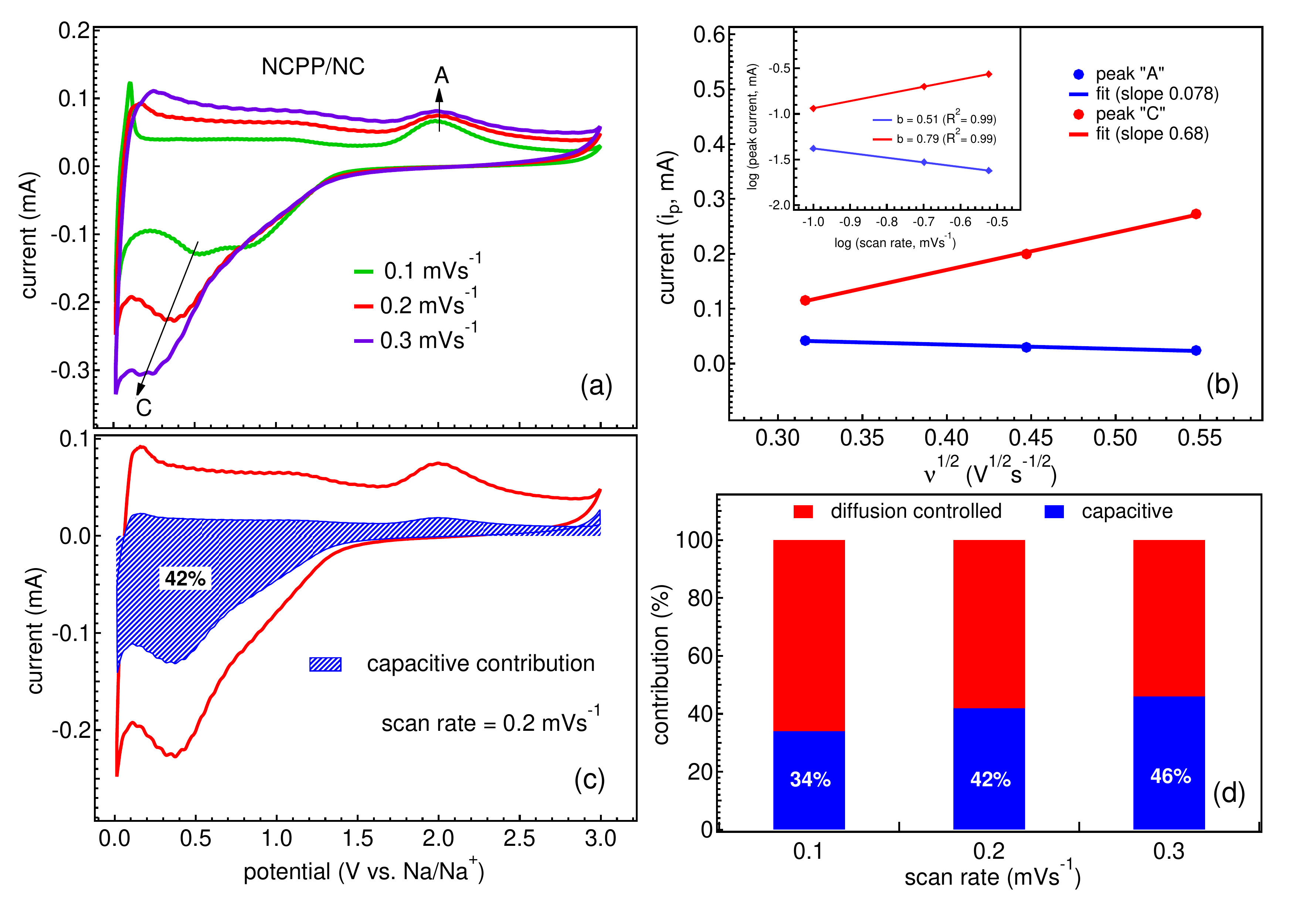}
\caption{(a) Cyclic voltammogram (third cycle) of NCPP/NC composite electrode at different scan rates, (b) the plot between peak current ($i_p$) with square root of scan rates for anodic and cathodic peaks for the NCPP/NC composite, inset in (b) shows the linear fit to log-log plot, (c) the extracted capacitive contributed area of 42\% at scan rate 0.2 mVs$^{-1}$, (d) the contribution of the capacitive and diffusive contributions in percentage at different scan rates for the NCPP/NC composite.} 
\label{EIS_DC}  
\end{figure*}

In order to understand the kinetics, electrochemical impedance spectroscopy measurements of the NCPP and NCPP/NC electrodes were performed at room temperature in the frequency range of 100~kHz to 0.01~Hz. As shown in Fig.~\ref{EIS_rate}(a), the Nyquist plots of the two electrodes consist of a depressed semicircle in the high-frequency region, which is attributed to the SEI film formation, contact resistance, and charge transfer resistance on the electrode/electrolyte interface. In addition, an inclined line in the low-frequency range corresponds to the sodium-diffusion process within the electrodes \cite{LiAdvmat2014}. We can calculate the ohmic resistance (R$_{\rm S}$), SEI contribution (R$_{\rm SEI}$) and charge transfer resistance (R$_{\rm ct}$) by fitting the circuit model, as shown in the inset of Fig.~\ref{EIS_rate}(a), which mimics the reaction mechanism into the electronic components. The R$_{\rm ct}$ resistance is observed to be lower for the NCPP/NC ($\approx$17~\ohm) electrode as compared to the pristine NCPP ($\approx$35~\ohm), which suggest the more conducting nature and shorten the diffusion path length to facilitate the Na$^{+}$ transference between electrode and electrolyte. Also, the contributions from SEI resistance and R$_{\rm S}$ are found to be around 1~\ohm~and 10~\ohm~for NCPP/NC, and 9~\ohm~and 19~\ohm~for NCPP, respectively. The reduced resistance values of the NCPP/NC electrode as compared to pristine NCPP are possibly due to N-doped carbon spheres, which decrease the mechanical strain and avoid volume expansion during the charge/discharge process. Additionally, the N doping enhances the electronic conductivity, providing continuous and rapid electron transport, which may accelerate the Na ion storage capacity of the electrode material. In Fig.~\ref{EIS_rate}(b) we show the comparative specific capacity performance of the NCPP and NCPP/NC electrodes at different current rates ranging from 0.5 to 5~C where 1~C = 170~mAg$^{-1}$. Evidently, the rate capability of the NCPP/NC composite is significantly better than that of the pristine NCPP. The NCPP/NC electrode delivers reversible capacity of 250, 173 and 112~mAhg$^{-1}$ at 0.5, 1 and  2~C current rates, respectively. At high current rate of 5~C, it delivers around 61~mAhg$^{-1}$ of discharge capacity and after reverting back to 0.5~C rate, the NCPP/NC electrode recovers around 129 mAhg$^{-1}$ of discharge capacity. The cycling stability of both the NCPP and NCPP/NC electrodes is tested at relatively high current rate of 1~C, as shown in Fig.~\ref{EIS_rate}(c). Here, we found that there is a fast capacity fading within first 50 cycles and thereafter the NCPP/NC electrode exhibits better cycle life even after 400 cycles where it shows around 44 mAhg$^{-1}$ specific capacity as compared to the 28 mAhg$^{-1}$ for pristine NCPP having 100\% Coulombic efficiency, see Fig.~\ref{EIS_rate}(c). However, the results in Figs.~\ref{EIS_rate}(b, c) suggest that initial low current rate may play crucial role for cycling stability and require detailed investigation in this context for further understanding to search new anode materials having open framework for sodium-ion batteries \cite{WangEA21}. Moreover, the EIS measurements of NCPP/NC electrode were carried out after 50, 100 and 200 charge-discharge cycles and compared in Fig.~\ref{EIS_rate}(d), where it is clearly visible that the charge transfer resistance along with Warburg resistance increase during the cycling. This is possibly due to small structural disorientation caused by cracks in the active material during cycling (discussed later). These observations are consistent as the specific capacity decays faster with the cycle numbers in Fig.~\ref{EIS_rate}(c). 

Note that the charge storage mechanism associated with any electrode can be interpreted from the kinetic analysis of CV measurements, which elucidate the contribution of surface controlled capacitive process and diffusion controlled redox process in the electrochemical reaction. In order to investigate these details for the NCPP/NC electrode, the cyclic voltametry measurements were performed at different scan rates ranging from 0.1 to 0.3~mVs$^{-1}$ and the results are shown in Fig.~\ref{EIS_DC}(a). With increasing the scan rate, the anodic peak `A' does not shift indicating good electrochemical reversibility; whereas, the cathodic peak `C' exhibit a shift towards lower potential convey a small polarization. These CV curves are used to define the relationship between the peak current ($i_p$) with scan rate ($v$) by the power law $i_p=$ a$v$$^{b}$ \cite{WangJMCA19}, where $b$ is an exponent. In Fig.~\ref{EIS_DC}(b), we plot the $i_p$ as a function of $v^{1/2}$ where the slope can be used to calculate the diffusion coefficient \cite{PuESM2019, LIUJMCA19}. The contribution of capacitive and diffusion process can be calculated by the value of exponent parameter $b$. It has been reported that if $b=$ 0.5, the charge storage mechanism refers to a diffusion-controlled electrochemical process and if $b=$ 1.0, it refers to an ideal capacitive controlled process \cite{WangJMCA19}. Further, we show the linear fit to the log-log plot, see inset of Fig.~\ref{EIS_DC}(b), to evaluate the $b$ values for oxidative and reductive peaks, which are found to be around 0.5 and 0.8, respectively. These values suggest for a combination of both diffusive (anodic regime) and pseudo-capacitive (cathodic regime) controlled Na$^+$ insertion/de-insertion into the NCPP/NC composite electrode \cite{ChenSm18, SunEA20, DongNJC18, ChenNC15}. 

The contributions of diffusive and capacitive processes are calculated using the equation, $i(\rm V)=$ $k_1v+k_2v^{1/2}$, where $i(\rm V)$ is the total current, $k_1v$ and $k_2v^{1/2}$ describe the capacitive part and diffusive controlled insertion, respectively. The linear fit to the $i({\rm V})/v^{1/2}$ versus $v^{1/2}$ for anodic and cathodic peaks give $k_1$ (slope) and $k_2$ (intercept), which are used to quantify the pseudo-capacitive contribution at different scan rates. Fig.~\ref{EIS_DC}(c) shows the capacitive contributed area of around 42\% at 0.2~mVs$^{-1}$. A higher diffusive contribution of 66\% is achieved at 0.1 mVs$^{-1}$, which decreases at higher scan rates due to more surface-controlled reactions. An increasing trend of capacitive behavior with 34\%, 42\% and 46\% contribution \cite{DongNJC18, ChenNC15} are observed at 0.1, 0.2 and 0.3 mVs$^{-1}$ scan rates. These results are presented in the percentage ratio versus the scan rate in Fig.~\ref{EIS_DC}(d), which indicate the pseudo-capacitive effect for the NCPP/NC composite electrode \cite{ChenSm18, SunEA20, DongNJC18, ChenNC15}. Further, the Na$^{+}$ ion diffusion coefficient of the electrode can be determined from the CV using the Randles-Sevcik equation \cite{PuESM2019, LIUJMCA19}:
\begin{eqnarray}
D_{\rm Na^+} = \frac{i_p}{2.69 \times 10^5n^{3/2}AC\nu^{1/2}}
\end{eqnarray}
where, the related parameters A is the surface area of the electrode (cm$^2$), C is the bulk concentration of the Na$^{+}$ ions (mol cm$^{-3}$), $n$ is the number of electrons involved in the redox reaction ($n$)~4, D$_{\rm Na^{+}}$ is the diffusion coefficient of sodium ions and $i_p$ corresponds to the peak current (in mA). Using eq.~1, the calculated diffusion coefficient of Na$^{+}$ for both the cathodic (C) and anodic (A) peaks is found to be around 1$\times$ 10$^{-10}$ cm$^2$s$^{-1}$, which is considered towards higher side for anode materials \cite{MaheshEA20}. 

\begin{figure}
\centering
\includegraphics[width=3.7in]{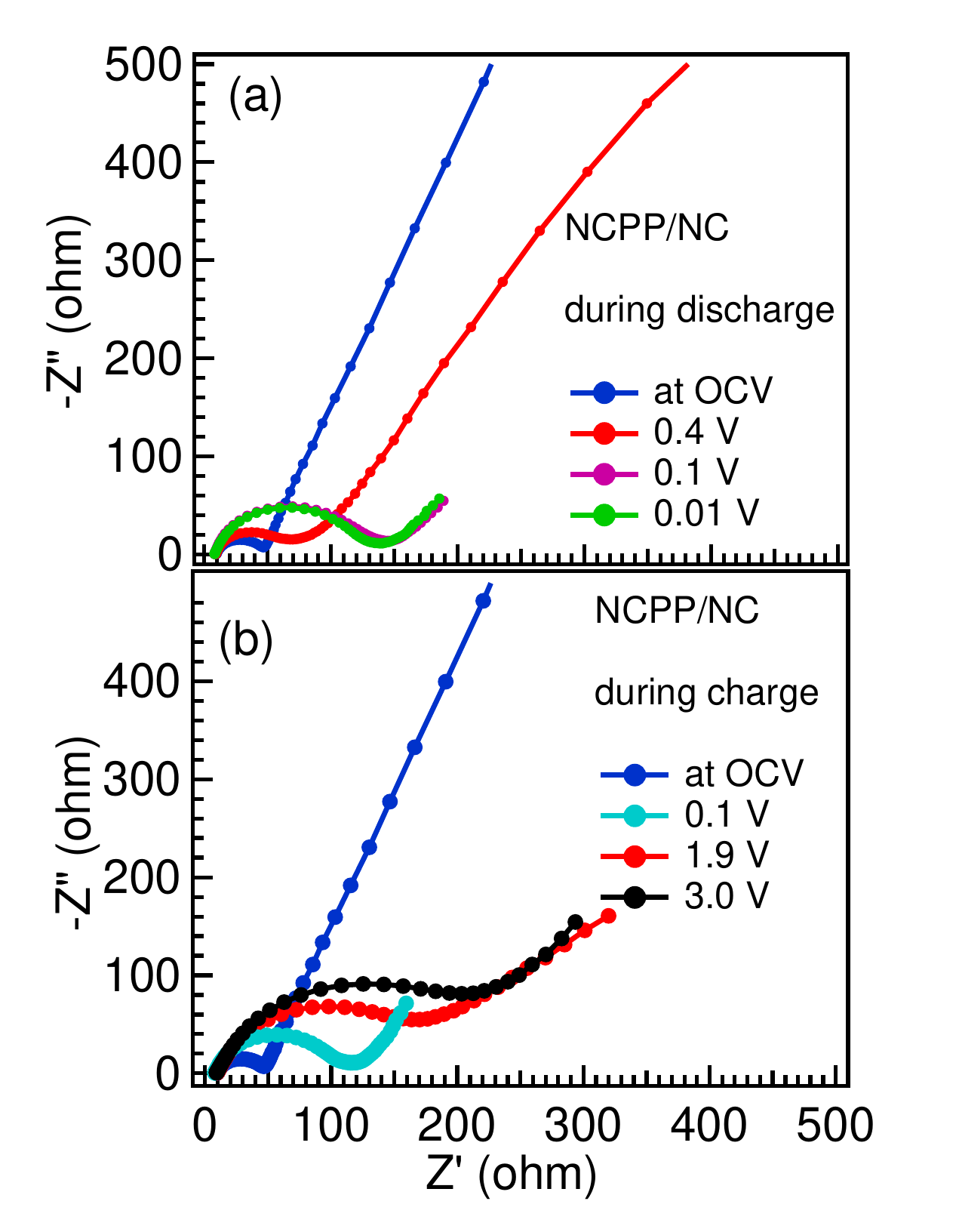}
\caption{Electrochemical impedance spectra (EIS) of the NCPP/NC composite anode measured at different voltages during (a) discharge process, and (b) charge process.}
\label{EIS_voltage}  
\end{figure}

\begin{figure}[h]
\centering
\includegraphics[width=3.45in]{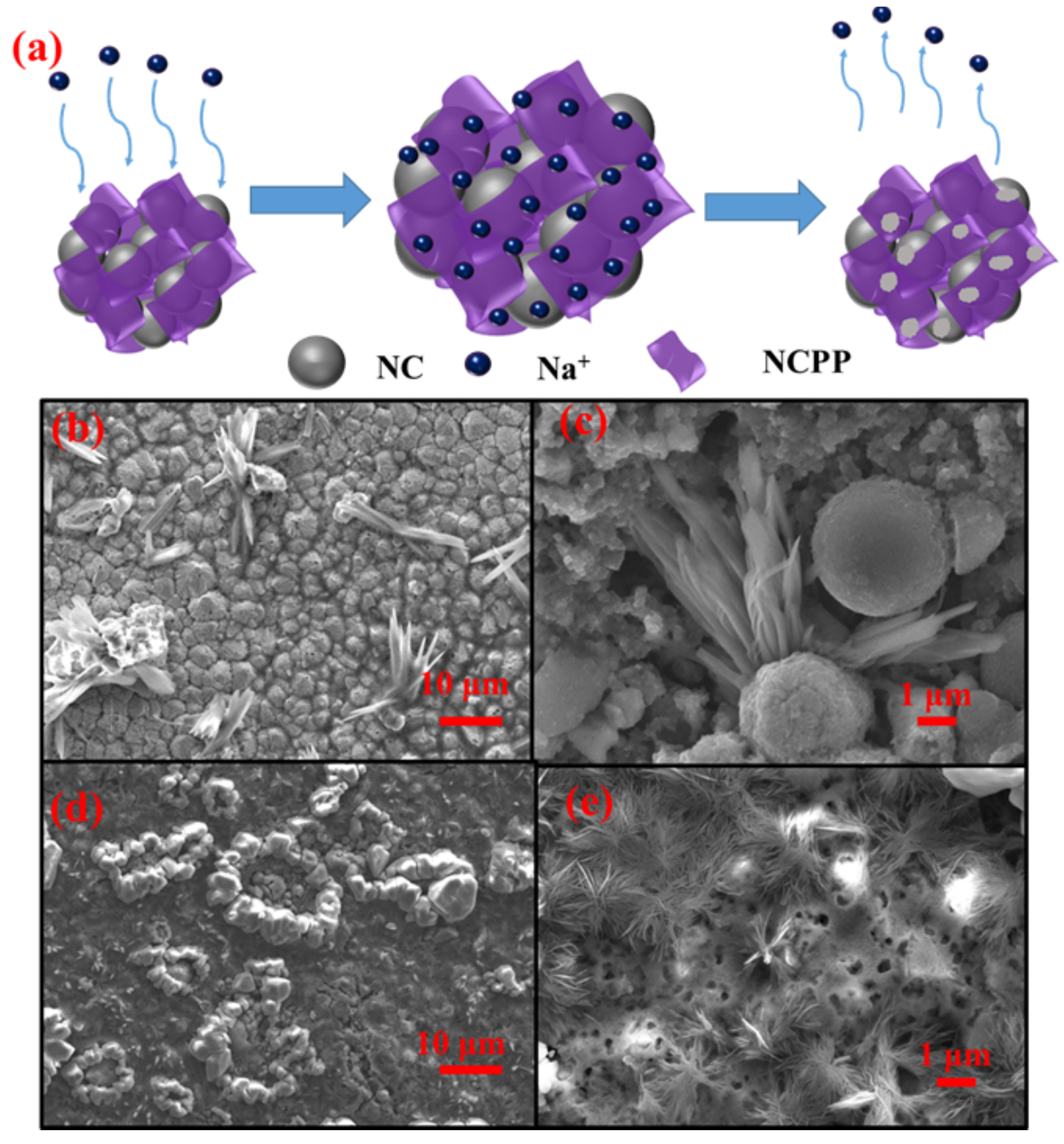}
\includegraphics[width=3.3in]{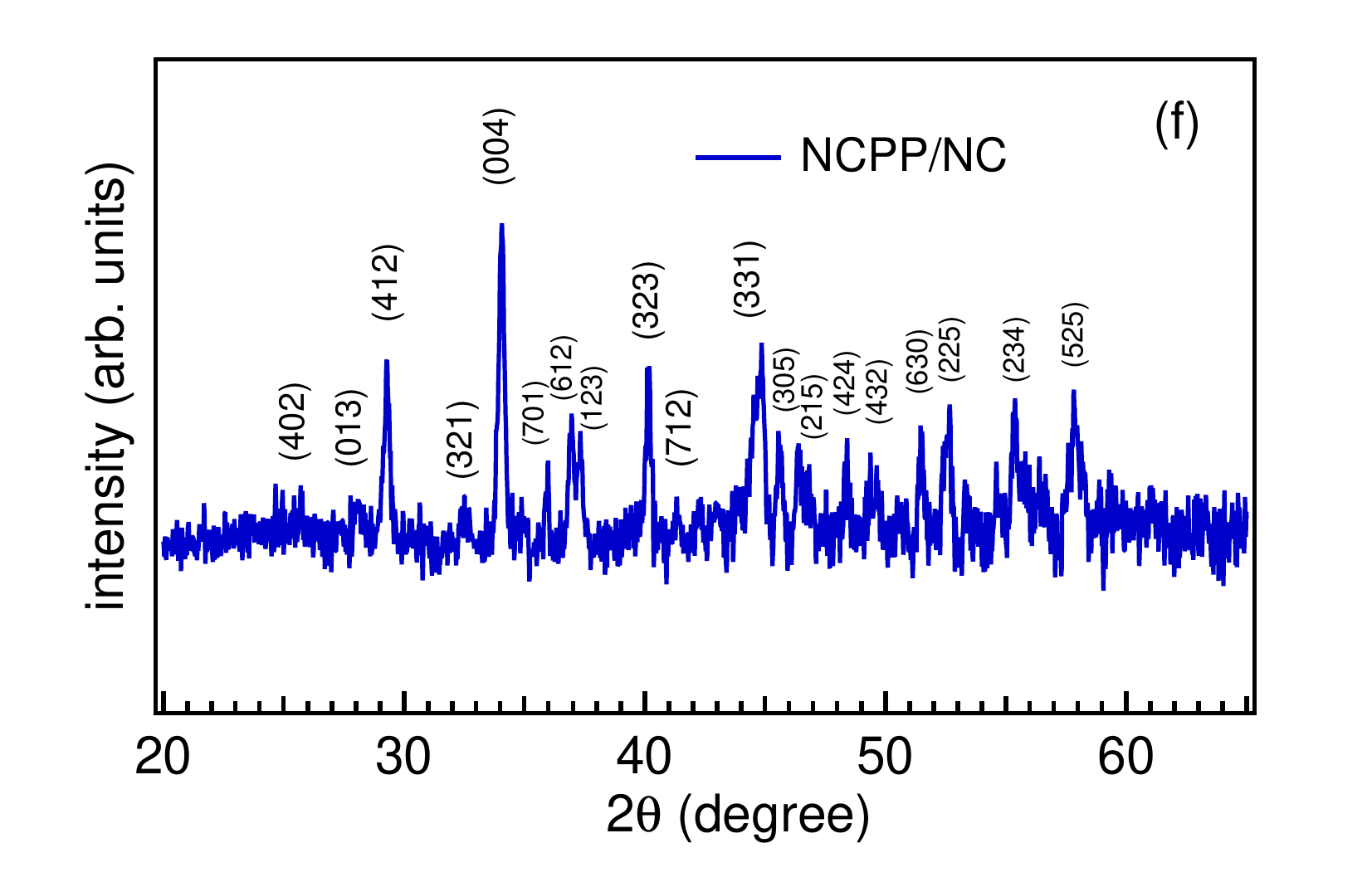}
\caption{(a) Schematic illustration of Na insertion and deinsertion into the NCPP/NC electrode; the FE-SEM images of the NCPP/NC electrode (b, c) fresh cell and (d, e) after 400 cycles of charge-discharge at 1~C rate, (f) the XRD pattern of the NCPP/NC electrode after 400 cycles of charge-discharge.} 
\label{SEM_XRD_cycle}  
\end{figure}

In order to examine the electrode kinetics, the EIS measurements were performed, which show the change in resistance at different states during first cycle of charge-discharge cycle for the NCPP/NC electrode, see Figs.~\ref{EIS_voltage}(a, b). Interestingly, the charge transfer resistance of 40~\ohm~is observed at OCV (2.0~V). During the discharge from 2.0 to 0.4~V and then to 0.1~V the R$_{ct}$ value is increased to around 55 and 135~\ohm, respectively and on completion of the discharge process at 0.01~V the R$_{ct}$ value decreased slight to 125~\ohm, see Fig.~\ref{EIS_voltage}(a). It is important to understand that during the discharging process, the Na ion insertion takes place via the infiltration of electrolyte to solid/electrolyte interface and entering to the crystal lattice of the NCPP/NC electrode. This process demands to cross a high kinetics hurdle which increased the R$_{ct}$ value. At discharge up to lower voltage around 0.1~V, the activation of the NCPP/NC electrode with formation of low valance Co species help to improve the electronic conductivity by reducing the charge transfer resistance and on complete discharge it exhibits a certain value of R$_{ct}$. Further, the R$_{ct}$ value increases during charging process, i.e., 120, 165, 200~\ohm~at 0.1, 1.9 and 3.0~V, respectively, see Fig.~\ref{EIS_voltage}(b). This is due to the slow reaction kinetics of the Na ion transportation at charge plateau voltage of about 1.9~V and the the formation of Co$^{2+}$ ions on complete charge, which diminish the electronic conductivity of Co--based oxide materials \cite{ShuklaPRB18, AjayPRB20}. 

Finally, the morphological and structural variation in the NCPP/NC electrode have been investigated after 400 cycles of de/sodiation by FE-SEM and XRD measurements. Fig.~\ref{SEM_XRD_cycle}(a) represents the schematic illustration of Na insertion and deinsertion into the NCPP/NC electrode. The SEM images for the fresh electrode are shown in Figs.~\ref{SEM_XRD_cycle}(b, c), where there is no evidence of cracks and pulverization in electrode surface. However, if we see the images taken after 400 cycles shown in Figs.~\ref{SEM_XRD_cycle}(d, e), it is clearly visible that the flakes and spheres are completely cracked into parts and the smooth surface of the electrode is shattered. The cracks lead to disintegration and isolation of certain area of active material in the post-cycling stage, which may generate strain/internal resistance and consequently result in the capacity fading \cite{PalaniCS18, CaprazEm18}. This can be attributed to chemo-mechanical degradation of the electrode as the formation of cracks in the active electrode surface area are visible in {\it ex-situ} post-cycling morphology analysis in Figs.~\ref{SEM_XRD_cycle}(c, e). These observations are also consistent with the {\it in-situ} EIS  measurements presented above in Fig.~6. Moreover, we study the structural alteration of the electrode material after 400 continuous charge-discharge cycles at 1~C rate. The collected XRD pattern is displayed in Fig.~\ref{SEM_XRD_cycle}(f). Interestingly, the reflections peaks are clearly visible, which indicates the structural stability of NCPP/NC electrode even after going through 400 cycles of sodiation/desodiation process. It is worth noting here that like NASICON structured composite materials \cite{JianCC15, WangACSAMI16, WeiACSAMI19, WangJMCA15, ZhaoJMCA19, FangAEM16, QiuNE19}, our results demonstrate that the mixed polyanionic composite with N doped carbon spares (NCPP/NC) could be a potential candidate as a negative electrode for sodium-ion batteries.    

\section{\noindent ~Conclusions}

In conclusion, we have successfully synthesized the NCPP/NC composite with simple facile and scalable synthetic route (sol-gel method) and examined as negative electrode for sodium-ion batteries. Here, the NCPP sheet is surrounded the surface of N doped carbon micro spheres, which improved the electrical and ionic conductivity and enhanced the sodium storage capacity of the NCPP/NC anode as compared to the pristine NCPP anode. The NCPP/NC anode delivers a reversible discharge capacity of 250~mAhg$^{-1}$ at 0.5~C current rate corresponds to intercalation and de-intercalation of four sodium ions and found to be stable up to high rate of 5~C (61~mAhg$^{-1}$). However, when starting the cell directly at high current rate 1~C, the capacity fading is relatively fast within 50 cycles, but after that the NCPP/NC anode shows slightly better electrochemical performance with stable life up to 400 cycles and appreciable Coulombic efficiency ($\approx$100\%). The extracted value of diffusion coefficient is found to be around 10$^{-10}$ cm$^2$s$^{-1}$. We demonstrate that in the composite, the carbon spheres offers intimate contact between the NCPP sheet, hinders the aggregation and alleviates the volume expansion in the NCPP during sodium ion insertion and extraction process. Additionally, the N doping in carbon spheres expedite the kinetics of Na ion transportation over the electrode/electrolyte interface, which boost the rate performance at elevated current rate. Hence, the development of composites could improve the low conductivity problem of polyanionic materials to make them potential candidate for forthcoming sodium-ion battery devices.

\section{\noindent ~ Acknowledgement} 

PKD, SKS and JP thank the DST project (DST/TMD/MECSP/2K17/07), MHRD and UGC, respectively for the fellowship. The Authors acknowledge the financial support from DST through "DST-IIT Delhi Energy Storage Platform on Batteries" (project no. DST/TMD/MECSP/2K17/07). We thank IIT Delhi for providing research facilities for sample characterization (XRD and Raman spectrometer at the physics department, FE-SEM and HR-TEM at CRF). We acknowledge the financial support from SERB through project under the core research grant (file no.: CRG/2020/003436). 

\newpage

{\bf Author Contribution Statement} 

All the authors have contributed to this work including data analysis and manuscript preparation.

{\bf Notes} 

The authors declare no competing finencial interest.

\end{document}